\documentclass[11pt]{JHEP3}
\JHEPspecialurl{http://jhep.sissa.it/JOURNAL/JHEP3.tar.gz}

\usepackage{epsfig,multicol}
\usepackage{amsmath,amssymb}
\usepackage{mathrsfs}
\newcommand\fverb{\setbox\fverbbox=\hbox\bgroup\verb}
\newcommand\fverbdo{\egroup\medskip\noindent%
			\fbox{\unhbox\fverbbox}\ }
\newcommand\fverbit{\egroup\item[\fbox{\unhbox\fverbbox}]}
\newbox\fverbbox

\newcommand{\pslash}{p\kern-1ex /}
\newcommand{\qslash}{q\kern-1ex /}
\newcommand{\lslash}{l\kern-1ex /}
\newcommand{\sslash}{s\kern-1ex /}
\newcommand{\kaslash}{k_a\kern-2ex /}
\newcommand{\kbslash}{k_b\kern-2ex /}
\newcommand{\Dslash}{\mathcal{D}\kern-1.5ex /}

\newcommand{\beqa}{\begin{eqnarray}}
\newcommand{\eeqa}{\end{eqnarray}}

\newcommand{\ba}{\begin{eqnarray}}
\newcommand{\ea}{\end{eqnarray}}
\newcommand{\be}{\begin{equation}}

\newcommand{\alg}[1]{\mathfrak{#1}}

\title{Quasi-local formulation of the mirror TBA}

\author{J\'anos Balog and \'Arp\'ad Heged\H us\\
Research Institute for Particle and Nuclear Physics, 
Hungarian Academy of Sciences,
H-1525 Budapest 114, P.O.B. 49, Hungary\\}

\received{\today} 		%%
\accepted{\today}	
\abstract{
We present a method of removing all infinite sums from the 
various forms of the mirror TBA equations and the energy formula of the
AdS/CFT spectral problem. This new formulation of the TBA system is 
quasi-local because $Y$-functions that are connected by the TBA equations
are at most next to nearest neighbors with respect to the Y-system 
diagram of AdS/CFT.}

\begin{document}

%%%%%%%%%%%%%%%%%%%%%%%%%

\newcommand{\con}{\,\star\hspace{-3.7mm}\bigcirc\,}
\newcommand{\convu}{\,\star\hspace{-3.1mm}\bigcirc\,}
\newcommand{\Eps}{\Epsilon}
\newcommand{\gM}{\mathcal{M}}
\newcommand{\dD}{\mathcal{D}}
\newcommand{\gG}{\mathcal{G}}
\newcommand{\pa}{\partial}
\newcommand{\eps}{\epsilon}
\newcommand{\La}{\Lambda}
\newcommand{\De}{\Delta}
\newcommand{\nonu}{\nonumber}
\newcommand{\beq}{\begin{eqnarray}}
\newcommand{\eeq}{\end{eqnarray}}
\newcommand{\ka}{\kappa}
\newcommand{\ee}{\end{equation}}
\newcommand{\an}{\ensuremath{\alpha_0}}
\newcommand{\bn}{\ensuremath{\beta_0}}
\newcommand{\dn}{\ensuremath{\delta_0}}
\newcommand{\al}{\alpha}
\newcommand{\bm}{\begin{multline}}
\newcommand{\fm}{\end{multline}}
\newcommand{\de}{\delta}

%%%%%%%%%%%%%%%%%%%%%%%%%%%%%%%%%%%%%%%%%%%%%%%%%%%%%%
\section{Introduction}

One of the most important problems in testing the AdS/CFT correspondence \cite{adscft} 
in the planar limit is to determine the finite size spectrum of the $AdS_5 \times S^5$
superstring sigma model \cite{AFrev}. After integrability was discovered in the string
worldsheet theory, the mirror Thermodynamic Bethe Ansatz (TBA) technique was
proposed \cite{AJK,AF07} to determine nonperturbatively the spectrum of the string theory.
The TBA equations for AdS/CFT were first derived for the ground state 
\cite{AF09a,Bombardelli:2009ns,AF09b,GKKV09,AF09d} and
then using an analytic continuation trick~\cite{DT} TBA equations were conjectured for certain classes of states in the $\alg{sl}(2)$
sector of the theory \cite{GKV09b,Arutyunov:2009ax,BH-BJ}.

An important property of the TBA equations is that the unknown functions ($Y$-functions)
satisfy the so called Y-system functional relations \cite{GKV09}. Recently an alternative  derivation
of the TBA equations appeared in \cite{Tateo1} where it has been shown that the complicated 
ground state TBA equations can be derived from the Y-system if it is 
supplemented by discontinuity functional equations relating the square-root discontinuities
of the various Y-functions. Later the derivation has been extended to excited states \cite{BH09}, too.

The main advantage of the Y-system based approach to the TBA problem is that it gives a deeper 
insight into the analytic properties of the $Y$-functions and opens the way to understand
the analytic properties of the $T$ and $Q$ functions, which are the elementary building blocks
of the underlying Y-system. This knowledge is indispensable if we want to construct an NLIE 
system formulated in terms of finitely many unknown functions for the AdS/CFT spectral problem.

So far the proposed TBA equations satisfied all tests in both the weak and strong coupling limits.
Their correctness has been nicely demonstrated by the convincing agreement with gauge theory results 
\cite{Sieg,Vel,KL02,40_5} in the weak coupling limit \cite{AFS,BHxxx,BH-BJ} through
the generalized L\"uscher approach \cite{JL07,BJ08,Bajnok:2008qj,BJ09,Lukowski:2009ce}, and
with string theory results in the strong coupling limit 
\cite{GKV09b,Gromov,ujKazi,Frolnum,StrongKonishi}.

In spite of the success of TBA technique in AdS/CFT, it has some obvious disadvantages as well.
First of all, like all the TBA equations of known sigma-models it contains infinitely many unknown
functions, which makes the study of their properties, both analytically and numerically, 
difficult.
Secondly, contrary to the case of relativistic models, these TBA equations cannot be formulated 
in a local form, namely a $Y$-function associated to a node of the $Y$-diagram of the model 
cannot be expressed, in general, in terms of the $Y$-functions belonging to the neighboring 
nodes only.

This non-locality is related to the discontinuity structure of the $Y$-functions and the 
discontinuity relations \cite{Tateo1}. As a consequence of non-locality all 
formulations{\footnote  {Using the terminology of \cite{Arutyunov:2009ax} canonical, 
simplified and hybrid versions of the TBA equations.}}
of the mirror TBA equations presented so far
containes infinite sums of certain $Y$ combinations convoluted
with appropriate kernels on the right hand side of the equations.
If expressed in terms of $Y$-functions, the TBA energy formula has the same property. 

The purpose of this paper is to work out a method that enables us to replace all the 
infinite sums appearing in the TBA approach by expressions containing only
a few terms and depending on a few $Y$-functions only.

Here we focus our attention to the hybrid formulation of the equations \cite{Arutyunov:2009ax}. 
In this case all the infinite sums which appear 
in the TBA description (including the energy formula) are of the form:
\begin{equation} \label{sum0}
\sum\limits_{Q=1}^{\infty} L_Q \star {\cal K}_Q,
\end{equation}
where $L_Q=\ln(1+Y_Q)$ and ${\cal K}_Q \equiv {\cal K}_Q(u,v)$ is a kernel satisfying the relation:
\begin{equation}
{\cal K}_Q-s \star ({\cal K}_{Q-1}+{\cal K}_{Q+1})=\delta {\cal K}_Q, \qquad{\cal K}_0 \equiv 0, \qquad
Q=1,2,...
\label{kerid0}
\end{equation}

The fact that enables us to transform (\ref{sum0}) into a finite sum is that for the ${\cal K}_Q$ 
kernels appearing in the TBA equations $\delta {\cal K}_Q=0$, with the exception of a few values of the
index $Q$ only. See formulae (\ref{dKQszeruek})-(\ref{dKsl2}).

In this paper we will derive finite expressions for the infinite sums
\begin{equation} \label{sum1}
\Omega = \sum\limits_{Q=2}^{\infty} L_Q \star {\cal K}_Q.
\end{equation}

The simplification of $\Omega$ consists of two main steps.
First we extend the definition of the convolution for functions of two variables in 
a natural way. Next with the help of the simplified TBA equations we can derive a 
linear integral equation for $\Omega$, which can be solved in Fourier space giving our
final formula (\ref{final}). 

If we replace the infinite sums of type (\ref{sum1}) with our new finite formulas (\ref{final}) in the
TBA equations and in the energy formula, the resulting formulation of the TBA equations
becomes quasi-local{\footnote {Here quasi-locality means that at most next to nearest 
neighbor $Y$-functions are connected by the TBA equations.}} as well. 
We think that such a quasi-local formulation of the TBA equations not only
makes it possible to perform more accurate numerical simulations 
but it is also an important step towards the NLIE formulation of the planar 
AdS/CFT spectral problem. 

The paper is organized as follows: Section 2 contains the derivation of the finite expression
for $\Omega$. In section 3 we discuss the energy formula and evaluate the finite formula 
for $\Omega$ when all $Y$-functions are replaced by their asymptotic counterparts. 
In section 4 we discuss the kernels appearing the TBA equations and present the 
complete set of quasi-local mirror TBA equations. The paper is closed by our conclusions.

Throughout the paper we use the notations, conventions and definitions of \cite{Arutyunov:2009ax}.
In this short paper we cannot reproduce all the formulae giving the definition of 
the various kernels appearing in the TBA equations and thus unfortunately the paper is not
self-contained. The necessary definitions together with the list of the TBA equations 
in the form we are using them in this paper can also be found in \cite{BH09}.

\section{TBA sums}

Since the convolution operation between functions plays a 
central role in our considerations here we recall the relevant definitions. The 
objects appearing in these formulas are either \lq\lq functions'' (typically
the logarithm of some Y-function) or \lq\lq kernels''. \lq\lq Functions'' 
depend on one variable, kernels on two. Some kernels only depend on the 
difference of the two variables. This is the case for the most important kernel
$s(u,v)$:
\begin{equation}
s(u,v)=s(u-v),\qquad s(u)=\frac{g}{4\cosh\frac{\pi gu}{2}}.
\end{equation}
For the \lq\lq function'' $\star$ \lq\lq kernel'' and 
\lq\lq kernel'' $\star$ \lq\lq function'' type convolutions we have
\begin{equation}
(f\star F)(v)=\int_{-\infty}^\infty{\rm d}u\,f(u)F(u,v),\qquad\quad
(F\star f)(v)=\int_{-\infty}^\infty{\rm d}u\,F(v,u)f(u)
\end{equation}
and for \lq\lq kernel'' $\star$ \lq\lq kernel'' type convolutions
\begin{equation}
(F_1\star F_2)(u,v)=\int_{-\infty}^\infty{\rm d}w\,F_1(u,w)F_2(w,v).
\end{equation}
The modified convolutions
\begin{equation}
(f\,\hat\star\, F)(v)=\int_{-2}^2{\rm d}u\,f(u)F(u,v),\qquad
(f\,\check\star\, F)(v)=\left(\int_2^\infty+\int_{-\infty}^{-2}\right)
{\rm d}u\,f(u)F(u,v)
\end{equation}
(and similarly for $F_1\,\hat\star\, F_2$, $F_1\,\check\star\, F_2$)
are also used in some of the TBA integral equations. 

Note that the standard convolution\footnote{As is well known,
$f\star^{\rm old}g=g\star^{\rm old}f$ and $\star^{\rm old}$ reduces to 
ordinary product after Fourier transformation.} 
definition between two functions of one variable
\begin{equation}
(f\star^{\rm old}g)(u)=\int_{-\infty}^\infty {\rm d}w\,f(u-w)g(w)
\end{equation}
is not always the same as our convolution, but
\begin{equation}
f\star s=f\star^{\rm old}s,
\end{equation}
because $s$ is even.

In this paper we will use a new convolution definition, which always maps
two functions of two variables into a function of two variables. Functions of 
one variable will be treated as functions of two variables by writing
$f(u,v)\rightarrow f(u-v)$:
\begin{eqnarray}
(F_1\con F_2)(u,v)&=&\int_{\infty}^\infty{\rm d}w\,
F_1(u,w)F_2(w,v),\\
(f\con F)(u,v)&=&\int_{\infty}^\infty{\rm d}w\,
f(u-w)F(w,v),\\
(f\con g)(u,v)&=&\int_{\infty}^\infty{\rm d}w\,
f(u-w)g(w-v)=(f\star^{\rm old}g)(u-v).
\end{eqnarray}
The new convolution $\ \convu$ is an (in general) non-commutative, associative 
product of functions.

Finally we note that the prototype TBA equation
\begin{equation}
L=\Lambda\star s,
\end{equation}
where $L$ and $\Lambda$ are \lq\lq functions'', can be equivalently
written
\begin{equation}
\hat L=\hat\Lambda\con s,
\end{equation}
where
\begin{equation}
\hat f(u)=f(-u)
\end{equation}
for any \lq\lq function''.

The kernel identity (\ref{kerid0}) can be rewritten as
\begin{equation}
{\cal K}_Q-s \con ({\cal K}_{Q-1}+{\cal K}_{Q+1})=\delta {\cal K}_Q, 
\qquad{\cal K}_0 \equiv 0, \qquad
Q=1,2,...
\label{kerid}
\end{equation}
and we also introduce the notation $\overline{\cal K}_m=s\con{\cal K}_m$.
$\overline{\cal K}_m$ satisfies an identity similar to (\ref{kerid})
with $\delta\overline{\cal K}_m=s\con\delta{\cal K}_m$.

In this section we will present the simplification of the TBA sums
working, for simplicity, in the $\alg{sl}(2)$ sector of the 
theory\footnote{Performing
the calculation for the general case is straightforward.}.  
We will consider the infinite sums
\begin{equation}
{\cal A}=\sum_{Q=2}^\infty\,\hat L_Q\con{\cal K}_Q
\end{equation}
and
\begin{equation}
{\cal B}=\sum_{m=1}^\infty\,\hat{\cal L}_m\con\overline{\cal K}_{m+1}.
\end{equation}
After having computed ${\cal A}$ we can write our final result as
$\Omega(v)={\cal A}(0,v)$.

We recall that $L_Q=\ln(1+Y_Q)$ and (see ref. \cite{BH09})
\begin{equation}
{\cal L}_m=\ln\left(\tau_m\tau_{m+2}\left(1+\frac{1}{Y_{m\vert vw}}\right)
\right),\qquad m=1,2,\dots
\end{equation}
further $r_m=\ln(1+Y_{m\vert vw})\  m=1,2,\dots, \ r_0=0$ and we define analogously
\begin{equation}
{\cal R}_Q=\ln\left(\tau_Q^2\left(1+\frac{1}{Y_Q}\right)
\right),\qquad Q=2,3,\dots,\qquad{\cal R}_1=\ln\left(1+\frac{1}{Y_1}\right).
\end{equation}

Using the notation introduced above we write the simplified TBA integral 
equations (\ref{TBAmvw}) and (\ref{TBA1vw}) as \cite{BH09}:
\begin{equation}
{\cal L}_m=r_m+L_{m+1}\star s-(r_{m+1}+r_{m-1})\star s
-{\mathfrak f}_2\,\delta_{m,1}
\qquad m=1,2,\dots
\end{equation}
and equivalently
\begin{equation}
\hat{\cal L}_m=\hat r_m+\hat L_{m+1}\con s-
(\hat r_{m+1}+\hat r_{m-1})\con s-\hat{\mathfrak f}_2\,\delta_{m,1}
\qquad m=1,2,\dots
\label{TBAhat1}
\end{equation}
Here
\begin{equation}
{\mathfrak f}_2=H\,\hat\star\, s,\qquad H=\ln\left(\frac{1-Y_-}{1-Y_+}\right).
\end{equation}
Similarly from (\ref{TBAQ})
\begin{equation}
\hat L_Q=\hat{\cal R}_Q+2\hat{\cal L}_{Q-1}\con s-
(\hat{\cal R}_{Q+1}+\hat{\cal R}_{Q-1})\con s-\hat{\mathfrak f}_1\,\delta_{Q,2}
\qquad Q=2,3,\dots
\label{TBAhat2}
\end{equation}
where
\begin{equation}
{\mathfrak f}_1=\ln\tau_1^2\star s.
\end{equation}

Now we take the convolution of (\ref{TBAhat2}) with ${\cal K}_Q$ and sum 
over $Q$ from 2 to infinity. After rearranging terms and using the kernel 
identity (\ref{kerid}) we get
\begin{equation}
{\cal A}=2{\cal B}+\sum_{Q=2}^\infty\,\hat{\cal R}_Q\con\delta{\cal K}_Q+
\hat{\cal R}_2\con\overline{\cal K}_1-\hat{\cal R}_1\con\overline{\cal K}_2
-\hat{\mathfrak f}_1\con{\cal K}_2.
\label{eqA}
\end{equation}
After performing similar transformations we get from (\ref{TBAhat1})
\begin{equation}
{\cal B}=\sum_{m=1}^\infty\,\hat r_m\con\delta\overline{\cal K}_{m+1}
+\hat r_1\con\overline{\overline{\cal K}}_1+s\con s\con{\cal A}
-\hat{\mathfrak f}_2\con\overline{\cal K}_2.
\label{eqB}
\end{equation}
Eliminating ${\cal B}$ from (\ref{eqA}) and (\ref{eqB}) we get a linear 
integral equation of the form
\begin{equation}
{\cal A}-2s\con s\con{\cal A}=\omega
\label{lin}
\end{equation}
with
\begin{equation}
\begin{split}
\omega=\sum_{Q=2}^\infty\,\hat{\cal R}_Q\con\delta{\cal K}_Q+
\hat{\cal R}_2\con\overline{\cal K}_1-&\hat{\cal R}_1\con\overline{\cal K}_2
-\hat{\mathfrak f}_1\con{\cal K}_2\\
+2&\sum_{m=1}^\infty\,\hat r_m\con\delta\overline{\cal K}_{m+1}
+2\hat r_1\con\overline{\overline{\cal K}}_1-
2\hat{\mathfrak f}_2\con\overline{\cal K}_2.
\end{split}
\label{omega}
\end{equation}
The linear equation (\ref{lin}) can easily be solved since the second argument
of the unknown ${\cal A}(u,v)$ is a \lq\lq spectator'' only and the 
essential part of the problem can be formulated in terms of kernels depending 
on differences only and it can be algebraically solved in Fourier space. The 
solution is of the form
\begin{equation}
{\cal A}=M\con\omega
\end{equation}
with
\begin{equation}
M=\delta+2s_{1/2},\qquad s_{1/2}(u)=\frac{1}{2}s\left(\frac{u}{2}\right),
\end{equation}
where $\delta$ is the Dirac delta funcion. This solution further satisfies 
\begin{equation}
M\con s=\sigma_{1/2},\qquad \sigma_{1/2}=s_{1/2}^++s_{1/2}^-,\qquad
\sigma_{1/2}(u)=\frac{g}{2\sqrt{2}}\,\frac{\cosh\frac{\pi gu}{4}}
{\cosh\frac{\pi gu}{2}}.
\end{equation}
A further useful identity is
\begin{equation}
M\con s\con s=s_{1/2}.
\label{Mss}
\end{equation}
Putting everything together, ${\cal A}$ can be expressed explicitly:
\begin{equation}
\begin{split}
{\cal A}=&\sum_{Q=2}^\infty\,\hat{\cal R}_Q\con(\delta{\cal K}_Q+
2s_{1/2}\con\delta{\cal K}_Q)+\hat{\cal R}_2\con\sigma_{1/2}\con{\cal K}_1
-\hat{\cal R}_1\con\sigma_{1/2}\con{\cal K}_2\\
-&\ln\hat\tau_1^2\con\sigma_{1/2}\con{\cal K}_2
+2\sum_{m=1}^\infty\,\hat r_m\con\sigma_{1/2}\con\delta{\cal K}_{m+1}\\
&+2\hat r_1\con s_{1/2}\con{\cal K}_1-
2\hat{\mathfrak f}_2\con\sigma_{1/2}\con{\cal K}_2.
\end{split}
\end{equation}
In all applications we need in this paper the kernel functions satisfy
\begin{equation}
\delta{\cal K}_Q=0,\qquad Q\geq3
\end{equation}
and furthermore we are interested in $\Omega$ only and thus the original
convolutions can be used. Our final result is
\begin{equation}
\begin{split}
\Omega=&{\cal R}_2\star(\delta{\cal K}_2+
2s_{1/2}\star\delta{\cal K}_2)+{\cal R}_2\star\sigma_{1/2}\star{\cal K}_1
-{\cal R}_1\star\sigma_{1/2}\star{\cal K}_2\\
-&\ln\tau_1^2\star\sigma_{1/2}\star{\cal K}_2
+2r_1\star\sigma_{1/2}\star\delta{\cal K}_2
+2r_1\star s_{1/2}\star{\cal K}_1-2H\,\hat\star\, s_{1/2}\star{\cal K}_2.
\end{split}
\label{final}
\end{equation}
The identity (\ref{Mss}) was used to simplify the last term.

\section{Energy formula and asymptotic form of the identity}

One of the important applications of our method is the calculation of the
infinite sum
\begin{equation}
{\cal E}=-\frac{1}{2\pi}\sum_{Q=1}^\infty\,\int_{-\infty}^\infty{\rm d}u\,
L_Q(u)\,\frac{{\rm d}\tilde p^Q}{{\rm d}u}
\label{energy}
\end{equation}
occurring in the expression for the energy of the state 
\cite{Arutyunov:2009ax}.
We can apply our formalism to calculate this sum with
\begin{equation}
{\cal K}_Q(u,v)\Rightarrow-\frac{1}{2\pi}
\,\frac{{\rm d}\tilde p^Q}{{\rm d}u}=\frac{g}{2\pi}\left[
x^{[Q]\prime}(u)-x^{[-Q]\prime}(u)\right].
\end{equation}
This expression satisfies the kernel identity (\ref{kerid}), but approaches
constant limits\footnote{Note that $x(u)=(u-i\sqrt{4-u^2})/2$.}
for large $\vert u\vert$. This last property of this kernel
leads to a problem in applying (\ref{final}) since the functions ${\cal R}_1$,
${\cal R}_2$ behave as $\ln\vert u\vert$ for large $\vert u\vert$ and the
corresponding convolutions are divergent.

There are two possible solutions of this problem. First we can use the identity
\begin{equation}
x^{[Q]\prime}(u)-x^{[-Q]\prime}(u)=i\int_{\vert v\vert>2}{\rm d}v\,
K_Q(u-v)\,\frac{v+i\epsilon}{\sqrt{4-(v+i\epsilon)^2}},
\label{enid}
\end{equation}
where
\begin{equation}
K_Q(u)=\frac{Qg}{\pi}\,\frac{1}{Q^2+g^2u^2}
\end{equation}
and use our formalism for ${\cal K}_Q(u,v)\Rightarrow K_Q(u-v)$ to calculate
\begin{equation}
t_1=\sum_{Q=1}^\infty\,L_Q\star K_Q.
\end{equation}
Since $K_Q$ is well-behaved at infinity, all convolutions are now convergent
and after having computed $t_1(u)$ with our method we can calculate
(\ref{energy}) by multiplying it with $\frac{ig}{2\pi}\,\frac{u}{\sqrt{4-u^2}}$
and integrate the result for $\vert u\vert>2$ (slightly above the real line).

An alternative, second solution (which is also applicable to other cases of 
interest) is to consider the asymptotic limit\footnote{This will be indicated
by an upper index $^{(0)}$ on the corresponding function.}
of our functions. The asymptotic limit is obtained from the Bethe Ansatz 
and these asymptotic solutions satisfy TBA integral equations that can be 
obtained from (\ref{TBAmvw},\ref{TBA1vw},\ref{TBAQ}) by simply deleting all 
$(1+Y_Q)$ factors occurring on the right hand side of these equations \cite{GKV09}. 
In our notation the asymptotic TBA equations become
\begin{equation}
\hat{\cal L}^{(0)}_m=\hat r^{(0)}_m-
(\hat r^{(0)}_{m+1}+\hat r^{(0)}_{m-1})\con s-
\hat{\mathfrak f}^{(0)}_2\,\delta_{m,1}
\label{asTBAhat1}
\end{equation}
(for $m=1,2,\dots$) and
\begin{equation}
\hat L^{(0)}_Q=\hat{\cal R}^{(0)}_Q+2\hat{\cal L}^{(0)}_{Q-1}\con s-
(\hat{\cal R}^{(0)}_{Q+1}+\hat{\cal R}^{(0)}_{Q-1})\con s
+(\hat L^{(0)}_{Q+1}+\hat L^{(0)}_{Q-1})\con s
-\hat{\mathfrak f}^{(0)}_1\,\delta_{Q,2}
\label{asTBAhat2}
\end{equation}
(for $Q=2,3,\dots$)
Since the particle rapidities $u_j$ occurring in $\tau_1$ can be treated
simply as parameters, we will not distinguish between the asymptotic and
\lq\lq exact'' versions of the function ${\mathfrak f}_1$ and will use
\begin{equation}
{\mathfrak f}^{(0)}_1={\mathfrak f}_1=\ln\tau_1^2\star s.
\end{equation}

We can now apply the same steps to the asymptotic sums
\begin{equation}
{\cal A}^{(0)}=\sum_{Q=2}^\infty\,\hat L_Q^{(0)}\con{\cal K}_Q\qquad\qquad
{\cal B}^{(0)}=\sum_{m=1}^\infty\,\hat{\cal L}_m^{(0)}\con
\overline{\cal K}_{m+1}
\end{equation}
as we did for their \lq\lq exact'' counterparts. We find that ${\cal A}^{(0)}$
cannot be determined in this way (it cancels from the equations) but 
we obtain instead the following identity:

\begin{equation}
\begin{split}
0=&\sum_{Q=2}^\infty\,({\cal R}^{(0)}_Q-L^{(0)}_Q)
\star(\delta{\cal K}_Q+2s_{1/2}\star\delta{\cal K}_Q)
+({\cal R}^{(0)}_2-L^{(0)}_2)\star\sigma_{1/2}\star{\cal K}_1\\
-&({\cal R}^{(0)}_1-L^{(0)}_1)\star\sigma_{1/2}\star{\cal K}_2
-\ln\tau_1^2\star\sigma_{1/2}\star{\cal K}_2\\
+2&\sum_{m=1}^\infty\,r^{(0)}_m\star\sigma_{1/2}\star\delta{\cal K}_{m+1}
+2r^{(0)}_1\star s_{1/2}\star{\cal K}_1
-2H^{(0)}\,\hat\star\, s_{1/2}\star{\cal K}_2.
\end{split}
\label{asfinal}
\end{equation}

We can recognize that the asymptotic limit of the 
quantity defined by (\ref{omega}) appears here and by rearranging 
some of the terms it can be expressed as

\begin{equation}
\omega^{(0)}=L_2^{(0)} \con s \con {\cal K}_1
-L_1^{(0)} \con {\cal K}_1+\sum\limits_{Q=1}^{\infty} L_Q^{(0)}\con
\delta {\cal K}_Q.
\end{equation}
From this result it is clear that $\omega$, and using (\ref{lin}),
also $\Omega$ is exponentially small in the asymptotic limit, which
of course must be true but is not obvious from the result (\ref{final}).

Let us use the formula (\ref{asfinal})
for the calculation of the energy expression.
In this case we have $\delta K_Q=0$ for $Q\geq2$ and we note that if we
subtract it from (\ref{final}) the $\ln\tau_1^2$ term cancels from the 
final result and we obtain
\begin{equation}
\begin{split}
t_1=&L_1\star K_1+
({\cal R}_2-{\cal R}^{(0)}_2+L^{(0)}_2)\star\sigma_{1/2}\star K_1
-({\cal R}_1-{\cal R}^{(0)}_1+L^{(0)}_1)\star\sigma_{1/2}\star K_2\\
+&2(r_1-r^{(0)}_1)\star s_{1/2}\star K_1
-2(H-H^{(0)})\,\hat\star\, s_{1/2}\star K_2.
\end{split}
\label{ast1}
\end{equation}
Here the difference ${\cal R}_2-{\cal R}^{(0)}_2$ decays
fast enough for $\vert u\vert\to\infty$ and this fact and the similar 
behavior of the other terms allow us to apply the identity (\ref{enid}) 
backwards, term by term, in this formula. The final
expression for the energy sum (\ref{energy}) can be written as follows:
\begin{equation}
\begin{split}
{\cal E}=&L_1\star \tilde{J}_1+
({\cal R}_2-{\cal R}^{(0)}_2+L^{(0)}_2)\star\sigma_{1/2}\star \tilde{J}_1
-({\cal R}_1-{\cal R}^{(0)}_1+L^{(0)}_1)\star\sigma_{1/2}\star \tilde{J}_2\\
+&2(r_1-r^{(0)}_1)\star s_{1/2}\star \tilde{J}_1
-2(H-H^{(0)})\,\hat\star\, s_{1/2}\star \tilde{J}_2,
\end{split}
\label{finalenergy}
\end{equation}
where we introduced the notation 
$\tilde{J}_Q(u)=-\frac{1}{2 \, \pi}\frac{{\rm d}\tilde p^Q}{{\rm d}u}$.

\section{The quasi-local TBA equations}

In this section, for the sake of completeness, we present the quasi-local form of the 
$AdS_5 \times S^5$ mirror TBA equations.
The form of the first group of the equations follows from the Y-system relations thus they are 
local and their form is the the same as that of the simplified version of the equations. 
\begin{eqnarray}
Y^{(\alpha)}_{m\vert vw}&=&t^{(\alpha)}_{m\vert vw}\exp\left\{
\ln\left[\frac{(1+Y^{(\alpha)}_{m+1\vert vw})(1+Y^{(\alpha)}_{m-1\vert vw})}
{(1+Y_{m+1})}\right]\star s\right\},\qquad m\geq2,\label{TBAmvw}\\
Y^{(\alpha)}_{1\vert vw}&=&t^{(\alpha)}_{1\vert vw}\exp\left\{
\ln\left[\frac{(1+Y^{(\alpha)}_{2\vert vw})}{(1+Y_2)}\right]\star s
+\ln\left[\frac{1-Y_-^{(\alpha)}}{1-Y_+^{(\alpha)}}\right]\ \hat\star\ s
\right\},\label{TBA1vw}\\
Y^{(\alpha)}_{m\vert w}&=&t^{(\alpha)}_{m\vert w}\exp\left\{
\ln\left[(1+Y^{(\alpha)}_{m+1\vert w})(1+Y^{(\alpha)}_{m-1\vert w})\right]
\star s\right\},\qquad m\geq2,\label{TBAmw}\\\
Y^{(\alpha)}_{1\vert w}&=&t^{(\alpha)}_{1\vert w}\exp\left\{
\ln\left[1+Y^{(\alpha)}_{2\vert w}\right]\star s+ 
\ln\left[\frac{1-\frac{1}{Y_-^{(\alpha)}}}{1-\frac{1}{Y_+^{(\alpha)}}}\right]
\ \hat\star\ s \right\}\,,\label{TBA1w}\\
Y_Q&=&t_Q\exp\left\{\ln\left[
\frac{Y_{Q+1}\,Y_{Q-1}
(1+Y^{(+)}_{Q-1\vert vw})(1+Y^{(-)}_{Q-1\vert vw})}
{Y^{(+)}_{Q-1\vert vw}Y^{(-)}_{Q-1\vert vw}
(1+Y_{Q+1})(1+Y_{Q-1})}\right]\star s \right\},\quad Q\geq2.\label{TBAQ}
\end{eqnarray}
Here the source terms $t^{(\alpha)}_{...}$ correspond to the singularities of the
$Y$-functions within the physical strip (see below).
The second group of equations contains the quantity $\Omega$ which  is 
a functional of the vector of kernels ${\cal K}_Q$. To emphasize this fact from now on 
$\Omega$ in (\ref{sum1}) 
will be denoted by $\Omega({\cal K}_Q)$ to make this functional dependence explicit.
On the right hand side of the subsequent equations $\Omega({\cal K}_Q)$ always denotes 
the finite expression obtained for $\Omega({\cal K}_Q)$ in our final formula
(\ref{final}). 
The quasi-local group of equations take the form:
\begin{equation}
\frac{Y_-^{(\alpha)}}{Y_+^{(\alpha)}}=
\frac{R_pB_m}{B_pR_m}\exp\left\{-L_1 \star K_{1y} -\Omega(K_{Qy}) \right\},
\label{YmperYp}
\end{equation}
\begin{equation}
\begin{split}
Y^{(\alpha)}_+Y^{(\alpha)}_-=\left(t_-^{(\alpha)}\right)^2
&\exp\Bigg\{
2\ln\left[\frac{1+Y^{(\alpha)}_{1\vert vw}}{1+Y^{(\alpha)}_{1\vert w}}
\right]\star s
+ L_1\star\left[
-K_1+2K^{11}_{xv}\star s\right]  \\
& -\Omega(K_Q)+2 \, \Omega(K^{Q1}_{xv}\star s)
-\ln\left[
\left(\frac{R_p^+R_p^-}{R_m^+R_m^-}\right)^2
\frac{Q^{--}}{Q^{++}}\right]\star s\Bigg\},
\end{split}
\end{equation}
\begin{equation}
\begin{split}
\ln Y_1=&-L\tilde{\cal E}_1+f_1+
2r_1\star s\ \hat\star\ K_{y1}\\
&-2\ln\left[\frac{1-Y_-}{1-Y_+}\right]\ \hat\star\ s\star K^{11}_{vwx}
+2{\mathscr L}_-\ \hat\star\ K^{y1}_-+2{\mathscr L}_+\ \hat\star\ K^{y1}_+\\
&\qquad+ L_1\star K^{11}_{{\alg{sl}(2)}}
+\Omega(K^{Q1}_{{\alg{sl}(2)}})+2 \,\Omega(s\star K^{Q-1\,1}_{vwx}),
\end{split}
\label{hybrid}
\end{equation}
where 
\begin{equation}
f_1(u)=-\sum_{j=1}^N\ln S^{1*1}_{{\alg{sl}(2)}}(u_j,u)-2\left[T^\epsilon\star
K^{11}_{vwx}\right](u)\label{f1}\\
\nonumber
\end{equation}
with
\begin{equation}
T^{\pm\epsilon}(u)=\sum_{j=1}^N\ln t(u-u_j\pm i\epsilon), \qquad t(u)=\tanh\left( \frac{\pi g u}{4}\right).
\end{equation}
The source terms appearing in the TBA equations (\ref{TBAmvw}-\ref{hybrid}) contain the objects
that are subject to the quantization conditions. 
In the $\alg{sl}(2)$ sector they can be expressed in terms of the 
$\tau_m(u)=\prod\limits_{j=1}^{{\cal N}_{m}} 
t(u-\xi_{m,j})$ and $\tilde{\tau}_{m}(u)
=\prod\limits_{j=1}^{\tilde{\cal N}_{m}} t(u-\tilde\xi_{m,j})$ 
functions as 
follows:
\begin{eqnarray}
t_{m \vert vw}&=&\tau_m \tau_{m+2}, \qquad m=1,2,..., \label{tmvw}\\
t_{m \vert w}&=&\tilde{\tau}_m \tilde{\tau}_{m+2}, \qquad m=1,2,..., \label{tmw}\\
t_Q&=& \tau_Q^2, \qquad Q=2,3,..., \label{tQ}\\
t_1&=&1, \label{t1}\\
t_{-}&=&\tau_2 /\tilde{\tau}_2. \label{t-}
\end{eqnarray}
Here the $\xi_{m,j}$ and $\tilde{\xi}_{m,j}$ are the objects to be determined
by the quantization conditions:
\begin{equation}
1+Y^\pm_{m\vert vw}(\xi_{m+1,j})=0,\qquad m=1,2,\dots,\qquad
j=1,\dots,{\cal N}_{m+1},
\end{equation}
\begin{equation}
1+Y^\pm_{m\vert w}(\tilde\xi_{m+1,j})=0,\qquad m=1,2,\dots,\qquad
j=1,\dots,\tilde{\cal N}_{m+1}.
\end{equation}
The set $\{\xi_{1,j}\}_{j=1,...,{\cal N}_1}$ is equal to the set of physical rapidities 
$\{u_j\}_{j=1,...,N}$ determined by the exact Bethe equations:
\begin{equation}
1+Y_{1 (*)}(u_j)=0, \qquad j=1,..,N
\end{equation}
and the set $\{\tilde{\xi}_{1,j}\}_{j=1,...,{ \tilde{\cal N}}_1}$ is empty by definition.
Here $Y_{1(*)}$ denotes $Y_1$ analytically continued to the physical sheet. For a detailed explanation
see refs. \cite{Arutyunov:2009ax} and \cite{BH09}. 
 We note that because of the left-right symmetry of the $\alg{sl}(2)$ sector 
we have omitted the wing index $ ^{(\alpha)}$ in the formulae above.

In order to be able to use our result (\ref{final}) in the equations (\ref{TBAmvw}-\ref{hybrid}) 
the vector $\delta {\cal K}_Q$ associated to the kernels ${\cal K}_Q$
appearing in the mirror TBA problem must be known. 
Since in the expression (\ref{final}) 
for $\Omega({\cal K}_Q)$,  $\delta {\cal K}_Q$ with $Q \geq 2$ appears only, we  
list the results for those $\delta {\cal K}_Q$ entering the mirror TBA equations 
(\ref{TBAmvw}-\ref{hybrid}) for $Q \geq 2.$
They are given by:
\begin{eqnarray}
\delta K_{Q} &=& 0, \qquad \qquad \quad \delta K_{Qy} = 0,\qquad \qquad \delta K^{Q1}_{xv} = 0, 
\qquad \qquad Q\geq 2, 
\label{dKQszeruek} \\
\delta(s \star K^{Q-1, 1}_{vwx}) &=& \delta_{Q,2} \, \, s \, \star s \, \hat{\star} \, K_{y1}
, \qquad \qquad 
\delta K^{Q 1}_{\alg{sl}(2)} = - \delta_{Q,2} \, \, s, \qquad \qquad \ Q \geq 2. \label{dKsl2} 
\end{eqnarray}
Using (\ref{dKQszeruek}-\ref{dKsl2}) and (\ref{final}) all the $\Omega({\cal K}_Q)$ terms
can be explicitly evaluated. The substitution of these expressions into the 
equations (\ref{TBAmvw}-\ref{hybrid})
gives the quasi-local form of the mirror TBA equations.

\section{Conclusion}

In this paper we worked out a method that enabled us to remove all infinite sums from the 
mirror TBA equations and the energy formula of AdS/CFT 
and obtained a quasi-local formulation of the TBA problem. 
Quasi-locality means that in this form at most next to nearest
{neighbor}{\footnote {with respect to the Y-system diagram of AdS/CFT}} $Y$-functions are 
coupled by the TBA equations. 
  
We think that the quasi-local formulation of the mirror TBA equations is an important 
step towards the NLIE formulation of the AdS/CFT spectral problem since using it 
the application of techniques worked out for the relativistically invariant models 
\cite{DdV-SG,Dunning,SS-DdV,GKV08,KL10} becomes possible.
 
Indeed, the locality of the simplified version of the TBA equations %(\ref{TBAmw})% 
made it possible to find a hybrid-NLIE formulation \cite{RyoHybrid} of the AdS/CFT spectral problem, 
where the two $SU(2)$ wings of the mirror TBA equations were resummed reducing remarkably
the number of unknowns.

%%%%%%%%%%%%%%%%%%%%%%%%%%%%%%%%%%%%%%%%%%%%%%%%%%%%%%
%%%%%%%%%%%%%%%%%%%%%%%%%%%%%%%%%%%%%%%%%%%%%%%%%%%%%%

 \vspace{1cm}
{\tt Acknowledgements}

\noindent 
This investigation was supported by the Hungarian National Science Fund OTKA (under K 77400).

%\newpage

%%%%%%%%%%%%%%%%%%%%%%%%%

%%%%%%%%%%%%%%%%%%%%%%%%%

\end{document}